\documentclass[runningheads]{llncs}
\usepackage[utf8]{inputenc}
\usepackage[T1]{fontenc}
\usepackage[american]{babel}
\usepackage{amsmath,amsfonts,amssymb,mathtools}
\usepackage{booktabs}
\usepackage[obeyFinal]{todonotes}
\usepackage{cite,breakcites}
\usepackage[inline]{enumitem}
\usepackage{csquotes}
\usepackage{acronym}
\usepackage{xspace}
\usepackage{multirow}
\usepackage[super]{nth}
\usepackage{hhline} 
\usepackage{makecell}
\newcolumntype{L}{m{0.035\textwidth}<{\centering}} 
\usepackage{color}
\usepackage{colortbl}
\usepackage{float}
\usepackage{pifont} 

\PassOptionsToPackage{hyphens}{url}
\usepackage[colorlinks=false,
            breaklinks=true]{hyperref}
\usepackage{cleveref}

\floatstyle{boxed}
\newfloat{experiment}{ht}{ex}
\floatname{experiment}{Experiment}
\Crefname{experiment}{Experiment}{Experiments}

\newfloat{scheme}{ht}{sc}
\floatname{scheme}{Scheme}
\Crefname{scheme}{Scheme}{Schemes}

\spnewtheorem{threat}{Threat}{\bfseries}{\rmfamily}
\Crefname{threat}{Threat}{Threats}
\spnewtheorem{assumption}{Assumption}{\bfseries}{\rmfamily}
\Crefname{assumption}{Assumption}{Assumptions}

\newlist{threatdesc}{description}{1}
\setlist[threatdesc]{font=\normalfont\itshape}

\setlist{noitemsep}

\newcommand{\tb}[1]{\textbf{#1}}
\newcommand{\tbb}[1]{{\underline{\textbf{#1}}}}
\newcommand{\fullKraken}{Bro\tbb{k}e\tbb{r}age and M\tbb{a}r\tbb{k}et platform for p\tbb{e}rso\tbb{n}al data}
\newcommand{\xmark}{\ding{55}}%

\newtoggle{iscolor}
\toggletrue{iscolor}

\definecolor{LINblue}{rgb}{.12, .55, .81} 
\definecolor{LINgreen}{rgb}{.44, .68, .28} 
\definecolor{LINorange}{rgb}{1, .6, 0} 

\makeatletter
\newif\ifAC@uppercase@first%
\def\Aclp#1{\AC@uppercase@firsttrue\aclp{#1}\AC@uppercase@firstfalse}%
\def\AC@aclp#1{%
  \ifcsname fn@#1@PL\endcsname%
    \ifAC@uppercase@first%
      \expandafter\expandafter\expandafter\MakeUppercase\csname
fn@#1@PL\endcsname%
    \else%
      \csname fn@#1@PL\endcsname%
    \fi%
  \else%
    \AC@acl{#1}s%
  \fi%
}%
\def\Acp#1{\AC@uppercase@firsttrue\acp{#1}\AC@uppercase@firstfalse}%
\def\AC@acp#1{%
  \ifcsname fn@#1@PL\endcsname%
    \ifAC@uppercase@first%
      \expandafter\expandafter\expandafter\MakeUppercase\csname
fn@#1@PL\endcsname%
    \else%
      \csname fn@#1@PL\endcsname%
    \fi%
  \else%
    \AC@ac{#1}s%
  \fi%
}%
\def\Acfp#1{\AC@uppercase@firsttrue\acfp{#1}\AC@uppercase@firstfalse}%
\def\AC@acfp#1{%
  \ifcsname fn@#1@PL\endcsname%
    \ifAC@uppercase@first%
      \expandafter\expandafter\expandafter\MakeUppercase\csname
fn@#1@PL\endcsname%
    \else%
      \csname fn@#1@PL\endcsname%
    \fi%
  \else%
    \AC@acf{#1}s%
  \fi%
}%
\def\Acsp#1{\AC@uppercase@firsttrue\acsp{#1}\AC@uppercase@firstfalse}%
\def\AC@acsp#1{%
  \ifcsname fn@#1@PL\endcsname%
    \ifAC@uppercase@first%
      \expandafter\expandafter\expandafter\MakeUppercase\csname
fn@#1@PL\endcsname%
    \else%
      \csname fn@#1@PL\endcsname%
    \fi%
  \else%
    \AC@acs{#1}s%
  \fi%
}%
\edef\AC@uppercase@write{\string\ifAC@uppercase@first\string\expandafter\string\MakeUppercase\string\fi\space}%
\def\AC@acrodef#1[#2]#3{%
  \@bsphack%
  \protected@write\@auxout{}{%
    \string\newacro{#1}[#2]{\AC@uppercase@write #3}%
  }\@esphack%
}%
\def\Acl#1{\AC@uppercase@firsttrue\acl{#1}\AC@uppercase@firstfalse}
\def\Acf#1{\AC@uppercase@firsttrue\acf{#1}\AC@uppercase@firstfalse}
\def\Ac#1{\AC@uppercase@firsttrue\ac{#1}\AC@uppercase@firstfalse}
\def\Acs#1{\AC@uppercase@firsttrue\acs{#1}\AC@uppercase@firstfalse}
\robustify\Aclp
\robustify\Acfp
\robustify\Acp
\robustify\Acsp
\robustify\Acl
\robustify\Acf
\robustify\Ac
\robustify\Acs

\def\AC@@acro#1[#2]#3{%
  \ifAC@nolist%
  \else%
  \ifAC@printonlyused%
    \expandafter\ifx\csname acused@#1\endcsname\AC@used%
       \item[\protect\AC@hypertarget{#1}{\acsfont{#2}}] #3%
          \ifAC@withpage%
            \expandafter\ifx\csname r@acro:#1\endcsname\relax%
               \PackageInfo{acronym}{%
                 Acronym #1 used in text but not spelled out in
                 full in text}%
            \else%
               \dotfill\pageref{acro:#1}%
            \fi\\%
          \fi%
    \fi%
 \else%
    \item[\protect\AC@hypertarget{#1}{\acsfont{#2}}] #3%
 \fi%
 \fi%
 \begingroup
    \def\acroextra##1{}%
    \@bsphack
    \protected@write\@auxout{}%
       {\string\newacro{#1}[\string\AC@hyperlink{#1}{#2}]{\AC@uppercase@write
#3}}%
    \@esphack
  \endgroup}
\makeatother

\setlist*[enumerate]{label=(\arabic*)}

\newcommand{\mathcmd}[1]{\ensuremath{#1}\xspace} 
\newcommand{\prob}[2][]{\mathcmd{\Pr\left[{#2}\if!#1!\else\mid{#1}\fi\right]}} 

\newcommand{\gsk}{\mathsf{sk}} 
\newcommand{\pk}{\mathcmd{pk}} 

\newcommand{\usk}[1][]{\mathcmd{\mathit{usk}\if!#1!\else_{#1}\fi}} 
\usepackage{tikz,colortbl}
\usetikzlibrary{calc}
\usepackage{zref-savepos}

\newcounter{NoTableEntry}
\renewcommand*{\theNoTableEntry}{NTE-\the\value{NoTableEntry}}
  \newlength{\myl}%

\title{%
  Privacy-preserving Analytics for Data Markets using MPC%
  \thanks{This project leading to this publication has received funding from the European Union’s Horizon 2020 research and innovation programme under grant agreement No 871473 (``KRAKEN'').
This is the full version of a paper which appears in (15th) IFIP Summer School on Privacy and Identity Management (2020), Revised Selected Papers. \textcopyright\ Springer, 2020.
  }%
}

\author{Karl Koch\inst{1} \and
        Stephan Krenn\inst{2}\orcidID{0000-0003-2835-9093} \and
        Donato Pellegrino\inst{3} \and
        Sebastian Ramacher\inst{2}\orcidID{0000-0003-1957-3725}}
\institute{Graz University of Technology, Graz, Austria\\
           \email{karl.koch@iaik.tugraz.at} \and
           AIT Austrian Institute of Technology, Vienna, Austria\\
           \email{\{stephan.krenn,sebastian.ramacher\}@ait.ac.at}  \and
           TX Tomorrow Explored, Helsinki, Finland\\
           \email{donato@tx.company}}
\titlerunning{Privacy-preserving Analytics for Data Markets using MPC}
\authorrunning{K. Koch et al.}

\begin{document}

\acrodef{ssi}[SSI]{self-sovereign identity}
\acrodef{mpc}[MPC]{secure multi-party computation}
\acrodef{gdpr}[GDPR]{General Data Protection Regulation}
\acrodef{tls}[TLS]{transport-layer security}

\acrodef{ld}[]{LINDDUN}
\acrodef{dfd}[DFD]{data-flow diagram}
\acrodef{ua}[UA]{user action}

\maketitle

\begin{abstract}
  Data markets have the potential to foster new data-driven applications and help growing data-driven businesses.
  When building and deploying such markets in practice, regulations such as the European Union's General Data Protection Regulation (GDPR) impose constraints and restrictions on these markets especially when dealing with personal or privacy-sensitive data.

  In this paper, we present a candidate architecture for a privacy-preserving personal data market, relying on cryptographic primitives such as multi-party computation (MPC) capable of performing privacy-preserving computations on the data.
  Besides specifying the architecture of such a data market, we also present a privacy-risk analysis of the market following the LINDDUN methodology.
  
\end{abstract}

\section{Introduction}

For the last decades, the amount of data generated, processed, and shared has been ever-increasing~\cite{www:forbes-data-generation-trend}.
Especially personal data has become more and more interesting~\cite{www:data-trend}.
One of the trends in this area is fitness and health data as more and more people are using fitness trackers, e.g. Garmin's connect~\cite{www:garmin}, Apple's Health app~\cite{www:apple-health}, or Google's Fit app~\cite{www:google-fit}, where the collected data can then be used in clinical research~\cite{www:fitbit-for-health}.
Relatedly, machine learning-based approaches facilitate the development of small sensors for measuring bodily functions of chronically ill patients. 
As an example, diabetes patients needing to draw blood multiple times a day, benefit from novel, noninvasive monitoring methods of their blood sugar levels~\cite{DBLP:journals/js/GusevPSKKSASTT20,todd2017towards}.
Very recently, location data collected by mobile network operators have become of interest to help combat the COVID-19 pandemic~\cite{fix:journals/iacr/HelmingerKRW20,www:google-covid-tracking}.

However, personal data is highly sensitive and regulations such as the \ac{gdpr} have to be taken into account when collecting, transmitting, storing, or processing the data. Therefore, these regulations present unique challenges in the design of data markets offering any kind of personally identifiable data. To profit from the opportunities as an individual (e.g., by sharing fitness data with insurance companies for better premiums) or for the common good (e.g., for limiting the outbreak of pandemics or clinical research), these challenges have to be overcome first, as otherwise misuse of the data could lead to discrimination, e.g., on the job market or by insurance companies~\cite{www:insurance-personal-data}.

More and more data markets are tackling these challenges via cryptographic means offering a wide variety of differing privacy guarantees.
Mediacalchain~\cite{medicalchain} among many others facilitates the exchange of medical data end-to-end secured.
Agora~\cite{DBLP:journals/iacr/KoutsosPCTH20} goes a step further and offers a data market place for privacy-sensitive data built from functional encryption (FE)~\cite{DBLP:journals/cacm/BonehSW12} where data consumers are able to buy evaluations of certain functions on user data.
Thereby only the results of the evaluation are exchanged without the need to transfer the original data sets.
Besides proposing the specific FE-based architecture, Koutsos et al. also provide a security model for confidentiality of processed data and consider payments in their analysis.
However, the security model does not consider confidentiality of the data against the broker.
Similarly, MyHealthMyData~\cite{DBLP:conf/edbt/Morley-Fletcher17} offers FE-based analytics with the focus on medical data while providing confidentiality of the data throughout the system.
Enveil~\cite{enveil} is a platform for outsourced computation using fully homomorphic encryption (FHE), but also offers possibilities for consumers to perform some analytic functions on the data.
Wibson~\cite{DBLP:journals/corr/abs-2001-08832} provides a smart contract based market place focusing on different privacy aspects, namely the privacy of seller's and buyer's identity.


\paragraph{\fullKraken\ (KRAKEN).}
The H2020 project KRAKEN~\cite{www:kraken-h2020} develops a data market for privacy-sensitive data.
To achieve that, the project \textquote{aims to enable the sharing, brokerage, and trading of potentially sensitive personal data, by returning the control of this data to citizens (data providers) throughout the entire data lifecycle}~\cite{www:kraken-h2020}.
KRAKEN mainly builds upon three pillars:
\begin{enumerate*}[label=(\roman*)]
  \item a data market place,
  \item \ac{ssi}~\cite{DBLP:journals/corr/abs-1712-01767}, and
  \item a toolbox of cryptographic primitives for privacy-preserving computation
\end{enumerate*}
to achieve that.
The market place acts as a broker between data providers and consumers.
\ac{ssi} is used to manage authentication, authorization, and, e.g., key management between data consumer and producer.
Privacy-preserving cryptographic protocols and primitives including \ac{mpc} are used to enable privacy-preserving analytics.

\subsection{Contribution}
In this paper, we propose and analyse a candidate architecture for the KRAKEN personal data marketplace that provides privacy-preserving distributed analytics features through the usage of \ac{mpc}.
The KRAKEN platform ensures user privacy and security of the overall system by relying on the decentralization of its core subsystems, SSI-based user management, and \ac{mpc}-based processing of data.
KRAKEN does not provide user data to the buyers. The core goal is to link buyers and sellers on the basis of metadata and policies, and enable data transfer between them in a privacy-preserving and decentralized manner.
Thereby, KRAKEN closes the gap left open in Agora.
Our contribution is twofold:
 
  \textbf{Architecture.}
  We describe a candidate architecture of the KRAKEN platform in detail, thereby explaining the necessary cryptographic background, suggesting instantiations of the building blocks to be used, and justifying any necessary design choices.
	The platform is designed in a way that allows for computations over inputs from potentially many different data sources in a single computation to allow for, e.g., statistics over many users.

  \textbf{Privacy analysis.}
    To validate the privacy requirements of the architecture, data flow diagrams and a privacy analysis based on LINDDUN~\cite{DBLP:journals/re/DengWSPJ11} are presented.
    This analysis considers all the privacy goals of KRAKEN, defines the related threats, and proposes mitigation strategies.

\paragraph{Paper outline.}
The remainder of this document is structured as follows.
In Section~\ref{sec:preliminaries}, we describe the necessary background on cryptographic building blocks and the LINDDUN methodology.
Then, in Section~\ref{sec:architecture}, we propose our architecture for a privacy-preserving data market, for which we then give a in-depth privacy analysis in Section~\ref{sec:linddun}.
Finally, we briefly conclude and sketch possible future research directions in Section~\ref{sec:conclusion}.

\section{Preliminaries}\label{sec:preliminaries}
The following sections give the necessary background on cryptographic building blocks, self-sovereign identities, and the LINDDUN methodology.
\subsection{Cryptographic Building Blocks}
Besides standard primitives such as encryption, our cryptographic architecture relies on a set of advanced privacy-preserving cryptographic mechanisms, which we will briefly introduce in the following.
We want to stress that practically efficient solutions and instantiations are available for each of these building blocks.

\paragraph{Group signatures.}
  Group signatures, initially introduced by Chaum and van Heyst~\cite{DBLP:conf/eurocrypt/ChaumH91} allow a party to sign a message on behalf of a group.
  That is, the verifier receives cryptographic guarantees that a member of the group indeed signed the message, yet he does not learn the identity of the actual signer.
  To achieve this goal, a \emph{group manager} generates a group public key $gpk$ as well as a master secret key $msk$.
  Now, when a user $U$ joins the group, she engages in a protocol with the group manager to receive her secret key $sk_U$ which she uses for signing messages, while the verifier only needs access to $gpk$ to verify the validity of signatures.

  It is worth noting that in group signatures, signer privacy is not absolute:
  in order to avoid abuse, a dedicated \emph{inspector} holding a inspection secret key $isk$ is able to revoke anonymity and reveal the originator of a signature.
  For the remainder of this paper we assume that all inspector public keys are generated in a way that no entity knows the corresponding secret key (e.g., by setting the public key to the hash value of a public nonce), as the inspection feature is not required in our scenario.
  The resulting primitive is then akin to Intel's Enhanced Privacy ID (EPID) scheme~\cite{DBLP:conf/socialcom/BrickellL10}, which also allows for signing messages on behalf of a group without anonymity revocation functionality.
  Recently, Kim et al.~\cite{DBLP:journals/iacr/KimLAP20} proposed the first group signature scheme supporting batch verifications, which significantly speeds up the verification process in case of many signatures.

\paragraph{Zero-knowledge proofs of knowledge.}
  A zero-knowledge proof of knowledge (ZK\-PoK) is a two party protocol between a \emph{prover} and a \emph{verifier}, which achieves two intuitively contradictory goals:
  it allows the prover to convince the verifier that she knows a secret piece of information, while at the same time revealing no further information than what is already revealed by the claim itself.

  Such protocols were first introduced by Goldwasser et al.~\cite{DBLP:conf/stoc/GoldwasserMR85}, and are a central building block for many privacy-preserving applications such as anonymous credential systems~\cite{DBLP:conf/crypto/Chaum82,DBLP:conf/eurocrypt/CamenischL01}, voting schemes~\cite{DBLP:conf/acns/Groth05}, e-cash~\cite{DBLP:conf/scn/CamenischHL06}, or group signatures.
  In recent years, zero-knowledge succinct non-interactive arguments of knowledge (SNARKs)~\cite{DBLP:conf/innovations/BitanskyCCT12} achieving very high efficiency have gained significant attention.


\paragraph{Secure multi-party computation.}
  Since its introduction by Yao~\cite{DBLP:conf/focs/Yao82b}, secure multi-party computation (MPC) has developed to an important building block for a variety of privacy-preserving applications.
  It allows a group of nodes to jointly evaluate a function on their inputs without revealing the inputs to the other nodes or any trusted third entity.
  Two major branches of MPC exist:
  while techniques based on garbled circuits are more efficient for bitwise operations~\cite{DBLP:conf/focs/Yao82b}, integer arithmetic can be computed highly efficiently using secret sharing based mechanisms~\cite{DBLP:journals/ijisec/BogdanovNTW12,DBLP:journals/cacm/Shamir79} due to the algebraic properties of these schemes.
  Especially during the last decade, research has come up with practically efficient protocols which have also been deployed in real-world scenarios and products~\cite{sharemind,DBLP:journals/iacr/IonKNPSSSY17,duality}.

\subsection{Self-Sovereign Identity} The Self-Sovereign Identity (SSI)~\cite{DBLP:journals/corr/abs-1712-01767,DBLP:journals/csr/MuhleGGM18} model describes an identity management concept which grants the owners of digital identities complete control over their data.
The goal of SSIs are ensuring the security and privacy of users' identity data, full portability of the data, no central authorities, and data integrity.

\subsection{LINDDUN Methodology}
LINDDUN~\cite{DBLP:journals/re/DengWSPJ11} is a threat modeling methodology for systematically analyzing privacy threats in software architectures.
Threats are analyzed along the categories linkability, identifiability, non-repudiation, detectability, disclosure of information, unawareness, and non-compliance.
On top of the threat analysis, it offers mitigation strategies to handle the identified threats.
The analysis consists of the following, cf. also \Cref{fig:linddun}:
\begin{figure}[t]
    \centering
    \includegraphics[width=.6\textwidth]{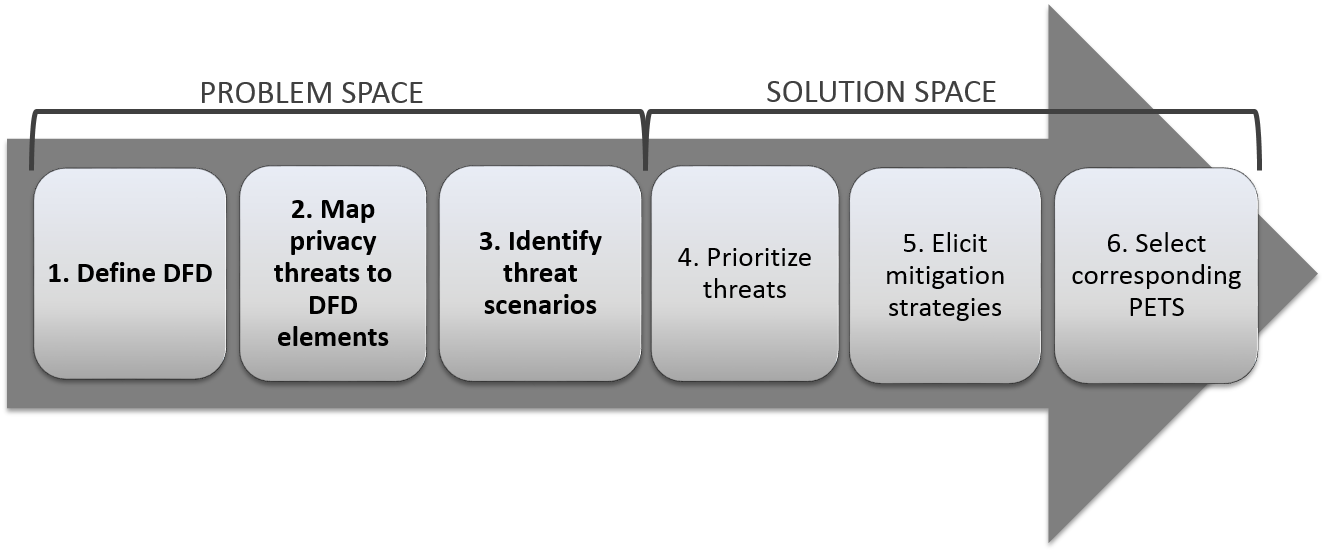}
    \caption{Overview of the \acl{ld} methodology~\cite{www:linddun}.}
    \label{fig:linddun}
\end{figure}
\begin{enumerate}
\item
  First, a \ac{dfd} is created detailing all involved entities, processes, trust boundaries, data stores and data flows.
\item
  In the second step, the \ac{dfd}s are mapped to the threat categories and for each element of the \ac{dfd} potential threat categories are identified.
\item
  Third, the identified threats are refined and documented. Also, assumptions that are made in the architecture are documented.
\item
  Fourth, the threats are prioritized based on their risk.
\item
  Next, mitigation strategies are defined, taking into account the risks that have been associated to each threat.
\item
  Finally, effective countermeasures are selected by mapping the defined mitigation strategies to suitable privacy-enhancing technologies.
\end{enumerate}

\section{KRAKEN Architecture}\label{sec:architecture}
As discussed by Koutsos et al.~\cite{DBLP:journals/iacr/KoutsosPCTH20}, a data market has to at least satisfy data privacy and output verifiability which are defined as follows:
\begin{description}
  \item[Data privacy:] No party can learn any information about on the data of the data owners. Only the result of computations on the data is known to the data consumers.\footnote{Note that in contrast to our definition, Agora allows data brokers, i.e. the market place, to learn the results as well.}
  \item[Output verifiability:] Authenticity of the data and results has to be ensured, i.e., falsified or incorrect results cannot be sold to a consumer.
\end{description}
Note that in contrast to Koutsos et al.~\cite{DBLP:journals/iacr/KoutsosPCTH20} we omit \emph{atomicity of payments}, which requires that data owners are correctly reimbursed, as for simplicity we do not (yet) consider per-access payments at this stage, but assume that data owners are compensated through a lump sum when uploading their data.
However, we consider it important that data owners stay in control of their data, and therefore request that data owners need to be able to define for which (types of) computations their data may be used, and which computations are considered to privacy-invasive by the data owner.

\medskip

\begin{figure}[t]
\centering
\includegraphics[width=\textwidth]{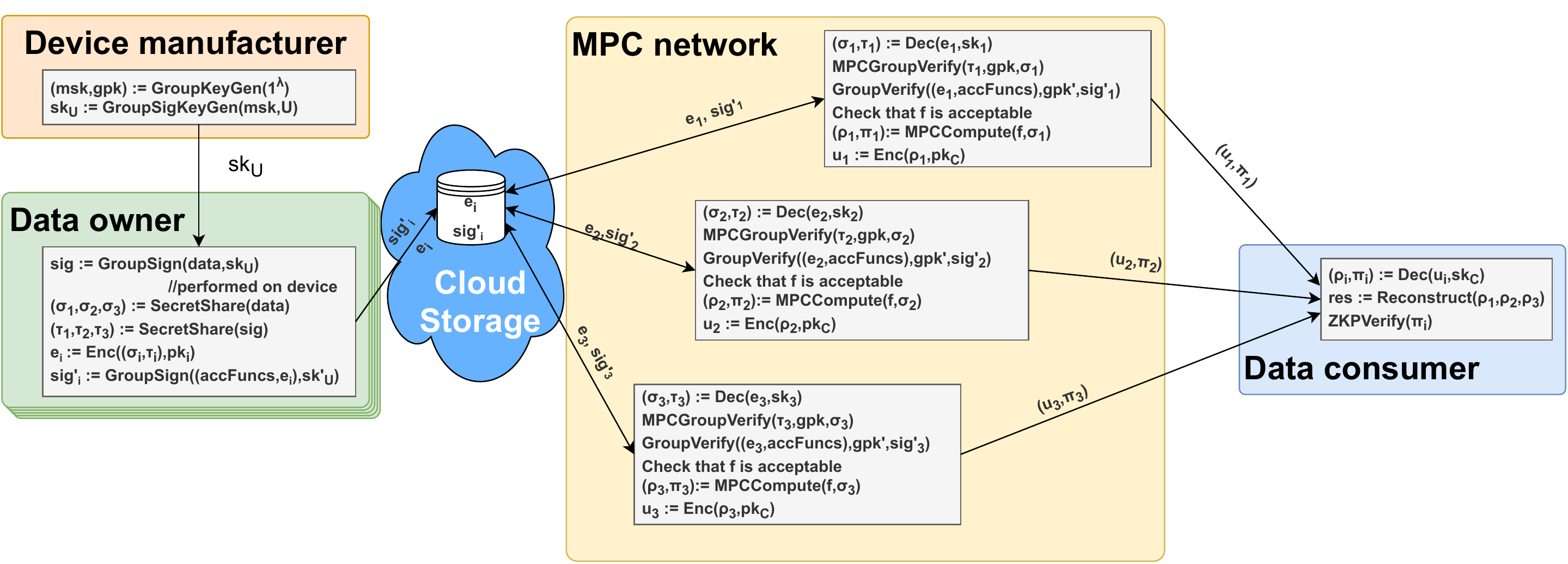}
\caption{KRAKEN cryptographic architecture overview.}
\label{fig:arch-overview}
\end{figure}

We will discuss one a candidate architecture of the KRAKEN market place.
The core idea of this architecture is centered around the use of MPC for privacy-preserving computation on data, group signatures to ensure data authenticity while preserving anonymity, and SNARKs for output verifiability.
In our architecture we consider the following actors:
\begin{description}
  \item[Device manufacturers] produce devices that collects sensitive data.
    All devices contain a signing key for a group signature scheme where the group is managed by the manufacturer.
    Devices sign the data using their key.
  \item[Data owners] use devices from the device manufacturer to collect sensitive data.
    The owner defines a family of functions which are allowed for performing computation on the data.
  \item[Data consumers] define the functions that should be used for the analysis of the data.
    They are in possession of a public-key encryption key.
  \item[Computation nodes] perform the analysis as defined by the data consumer on the data of the data owner using a secret sharing-based MPC protocol.
    They are in possession of a public-key encryption key.
    These computation nodes can be run by cloud computing providers that offer ``computation as a service'', various participants of the market place including the data owners and consumers, or by operator of the market place.
  \item[KRAKEN market place] handles the registration of data owners and consumers and manages listings of available data sets. The information is stored on an internal blockchain and database.
  \item[Cloud storage provider] offers storage to data owners without the need for registration.
\end{description}
An overview of the data processing is depicted in \Cref{fig:arch-overview}. We will now discuss the three typical data flows.

\paragraph{User registration.}
We start with the registration of a data owner and data consumers on the market place which is centered around credentials from a SSI system.
In this step we assume that they obtained their credential from an identity provider before, as this is a step only performed once independent of the registration at the market place.
\begin{enumerate}
  \item The user is in possession of SSI credentials and uses them to create an account on the data market place.
  \item The user and the market place perform the group signature joining procedure. At the end of the interaction, the user obtains a group signing key $\gsk_U'$ that she may use to sign policies specifying the types of computations that may be performed on her data.
\end{enumerate}

\paragraph{Data pre-processing and registration.}
After a data owner registered on the market place, she is able to register her data for subsequent analysis in the market place.
To do so, the data owner performs the following steps:
\begin{enumerate}
  \item The data owner collects data produced by her device, which is signed also by the device's group signature signing key $\gsk_U$.
  \item The data owner produces shares of the data and associated signature, using a secret sharing scheme compatible with the deployed MPC protocol, resulting in shares $\sigma_i$, $i=1,2,3$ -- one per MPC node.
  \item Using the public key $\pk_i$ of MPC node $i$, the data owner encrypts $\sigma_i$.
  \item The data owner signs the encrypted shares and an acceptable family of functions using their group signature key $\gsk_U'$.
  \item The data owner sends all information, i.e. encrypted shares, acceptable family of functions, and the signature under $\gsk_U'$, to a cloud storage provider of their choice.
  \item The data owner registers the offering on the market place and informs the market place on the location of the encrypted secret shares.
\end{enumerate}
For simplicity, in the proposed architecture, we assume that data owners are reimbursed by the market place via a lump sum when registering their data, and no further payments will take place on a per-usage basis.

\paragraph{Analysis.}
A consumer uses the market place to find data sets that are of interest and negotiates their use via the market place.
During the negotiation, the consumer declares the function to be evaluated on the data. Once an agreement is reached, the evaluation is performed as follows:
\begin{enumerate}
  \item The market place informs the computation nodes of the function to perform and the location of the data items.
  \item The computation nodes fetch the encrypted secret shares and signatures from the cloud storage.
  \item After receiving all encrypted shares and signatures and the function $f$, the computation nodes first verify, for each data item, the signature on the data evaluation policy, and that $f$ is an eligible function with respect to this policy.
  They then decrypt the shares, jointly verify the shared signatures, and start the MPC protocol to compute $f$ on the data.
  \item The shares obtained as result of the computation, are then encrypted with respect to the consumer's public key, $\pk_C$.
  \item The nodes provide a ZK-PoK/SNARK that they computed the function $f$ on the received data and that the obtained signature verified on the inputs.
  \item The nodes send the encrypted results and the proof to the consumer.
  \item The consumer decrypts the shares of the result and combines them to obtain the result of the evaluation. The consumer also verifies its correctness by checking the proofs sent by the nodes.
\end{enumerate}
Similarly to before, in the current architecture we assume that consumers pay the market place, but no (direct or indirect) payment from the consumer to the data owner takes place.
Further suggestions for a more fine-granular reimbursement concept can be found in Section~\ref{sec:conclusion}.

\section{\acl{ld} Analysis of KRAKEN}\label{sec:linddun}

We now present the LINDDUN analysis for KRAKEN's architecture.
In the analysis, we focus on three user actions:
\begin{enumerate*}
    \item Register user,
    \item Register availability of data, and
    \item Perform data analysis.
\end{enumerate*}

\begin{figure}[t]
    \centering
    \includegraphics[width=\textwidth]{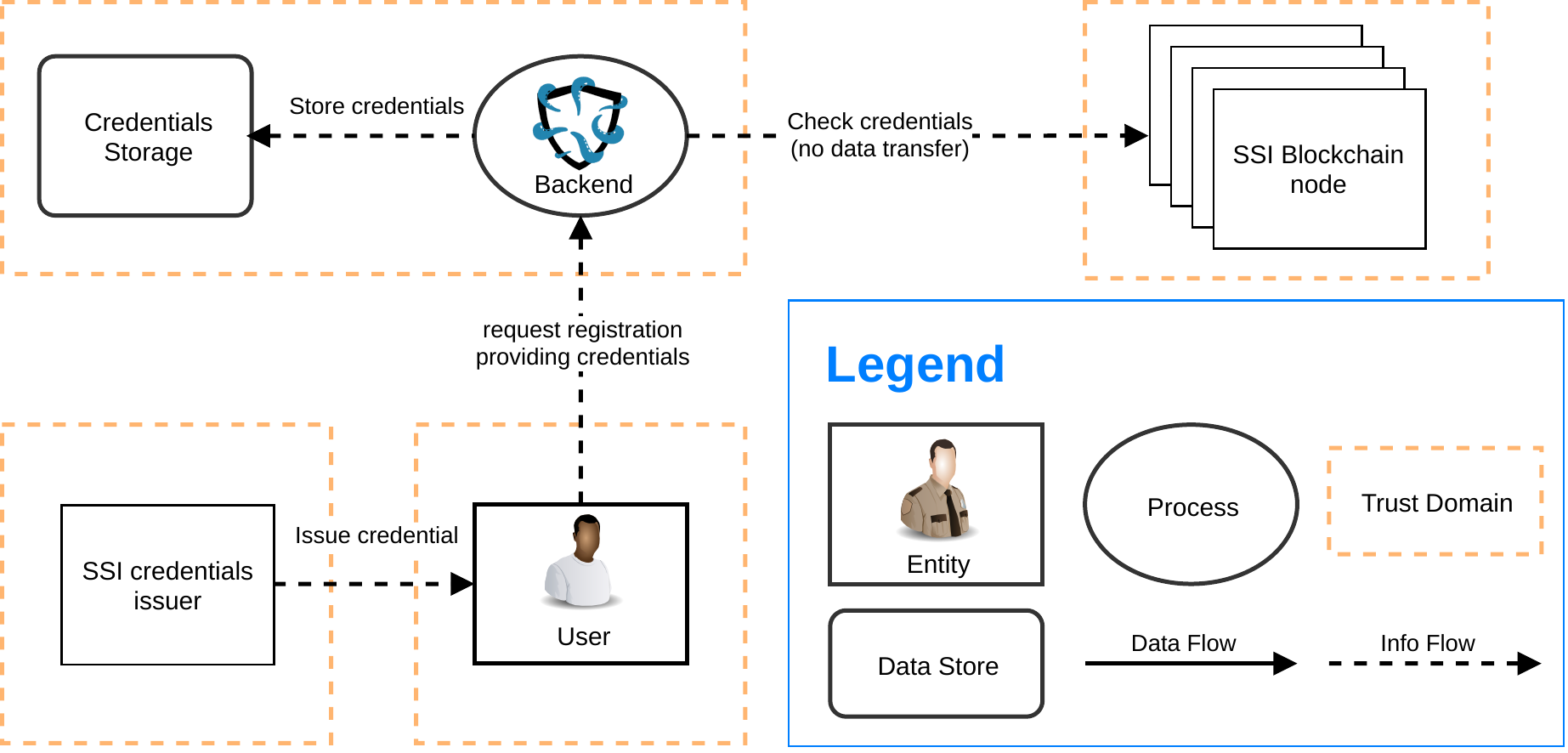}
    \caption{\ac{dfd} for the user action of registering.}
    \label{fig:lind:dfd:user-reg}
\end{figure}

\begin{figure}[t]
    \centering
    \includegraphics[width=1.3\textwidth]{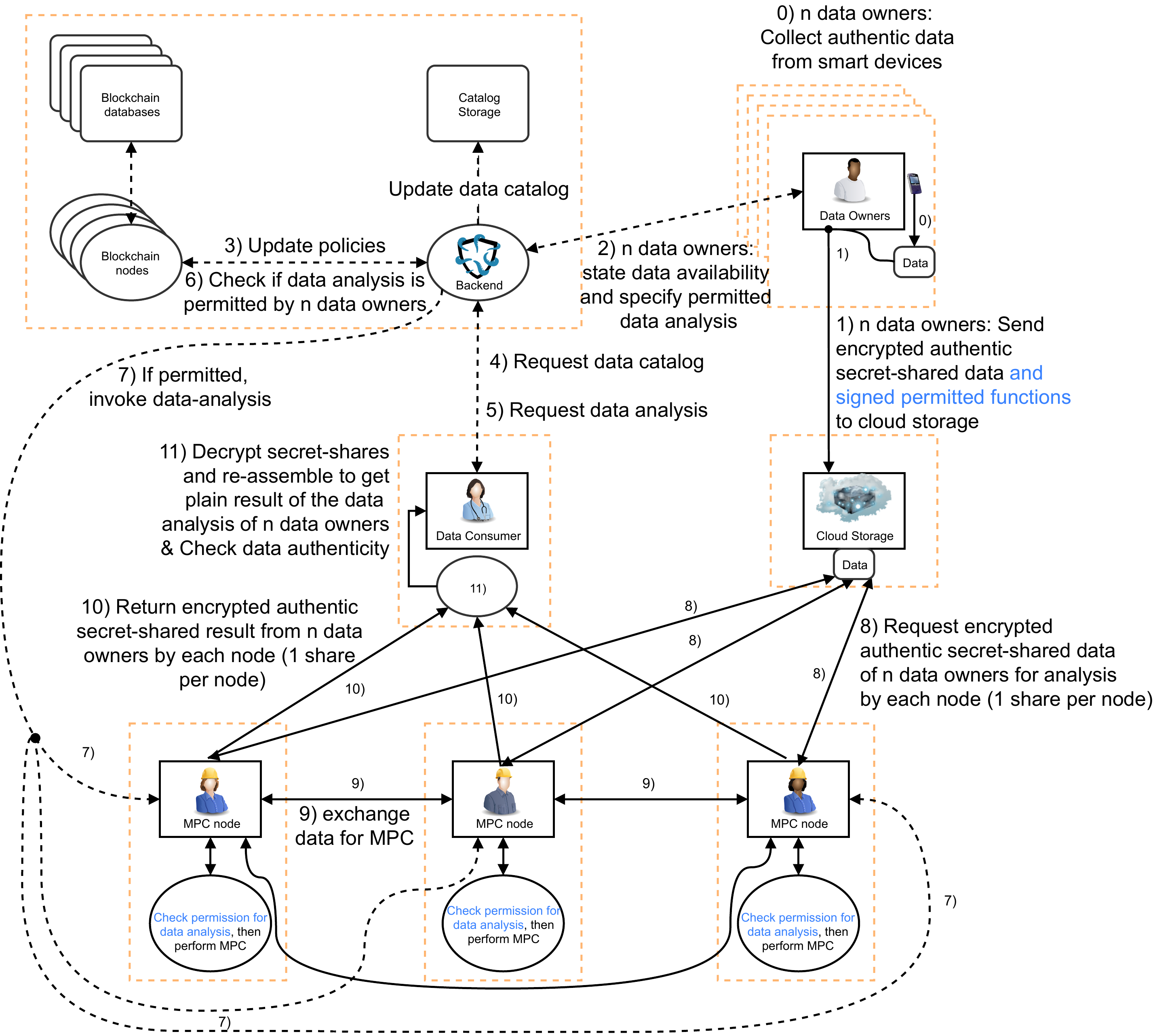}
    \caption{\ac{dfd} for the user actions of registering data and performing data analysis. The legend is as in Fig.~\ref{fig:lind:dfd:user-reg}.}
    \label{fig:lind:dfd:data-analysis}
\end{figure}

We start with the DFDs.
For reasons of clarity and comprehensibility, we split the data flow into two categories.
First, a \emph{real data flow}, which contains (parts of) personal data from data owners; such as encrypted shares or an analytics result.
Second, an \emph{info flow}, for other types of data flow; such as registering availability of data or invoicing data analysis.
\Cref{fig:lind:dfd:user-reg} visualizes the \ac{dfd} for the user action of registering a user.
\Cref{fig:lind:dfd:data-analysis} visualizes the \ac{dfd} for the user actions of registering availability of data and performing data analysis.

\subsection{Threat Tables}
We use the LINDDUN mapping template~\cite{DBLP:journals/re/DengWSPJ11} to map the DFD elements to the seven threat categories: Linkability, Identifiability,
Non-repudiation, Detectability, Disclosure of information, Unawareness, Non-compliance.
In the LINDDUN mapping template, entities are affected only by linkability, identifiability and unawareness threats, while the
other DFD elements are affected by every threat except unawareness.
We consider processes, flows and entities internal to trust domains not affected by any privacy threat from outside
due to the fourth assumption.
Data Flows and Info Flows are not affected by Non-compliance threat as the communications are secured with TLS.
The Non-compliance threat does not affect Data Stores as well as they are implemented with the data-minimization principle. 
Also processes are not affected by the Non-compliance threat as no sensitive information (user personal data) is handled by them;
in the case of MPC nodes the second assumption rules out the privacy threat.

\Cref{tab:lin:threat-table:user-registration} shows the threat table of our \nth{1} user action, performing user registration.
\begin{table}[t]
\centering
  \caption{\acl{ld}'s threat table of the \nth{1} user action, performing user registration.
            An \xmark\ in a cell indicates a privacy threat for the
            corresponding threat target. 
            Cells labeled by ``Ax'' are no threats because of the indicated assumptions.
            }
   \label{tab:lin:threat-table:user-registration}
  \begin{tabular}{ | m{.18\textwidth}<{\centering} | m{.3\textwidth}  || L | L | L | L | L | L | L | }
     \hline
     \multirow{2}{*}{\ac{dfd} Elements} & \multirow{2}{*}{Threat Target} & \multicolumn{7}{c|}{Privacy Threats} \\
                                                                           \cline{3-9}
                                        &                                & \tb{L} & \tb{I} & \tb{N}r & \tb{D}e & \tb{D}i & \tb{U} & \tb{N}c \\
    \hline \hline
    Data Store & Credentials Storage                      & \xmark & \xmark & & & & &  \\
    \hline
    \multirow{4}{*}{Info Flow}  & Issue credentials       & \xmark & \xmark & \xmark & \xmark &\cellcolor{\iftoggle{iscolor}{LINorange}{lightgray}}A3 & &  \\
                                \hhline{|~|--------|}
                                & Request Registration    & \xmark & \xmark & \xmark & \xmark &\cellcolor{\iftoggle{iscolor}{LINorange}{lightgray}}A3 & &  \\
                                \hhline{|~|--------|}
                                & Check credentials       & \xmark & \xmark & \xmark & \xmark &\cellcolor{\iftoggle{iscolor}{LINorange}{lightgray}}A3 & &  \\
                                \hhline{|~|--------|}
                                & Store credentials       & & & & & & &  \\
    \hline 
    Process & Backend                                     & & & & & & & \\
    \hline
    \multirow{3}{*}{Entity}     & User                    & & & & & & &  \\
                                \hhline{|~|--------|}
                                & SSI credentials issuer &\cellcolor{\iftoggle{iscolor}{LINgreen}{lightgray}}A1 &\cellcolor{\iftoggle{iscolor}{LINgreen}{lightgray}}A1 &\cellcolor{\iftoggle{iscolor}{LINgreen}{lightgray}}A1 &\cellcolor{\iftoggle{iscolor}{LINgreen}{lightgray}}A1 &\cellcolor{\iftoggle{iscolor}{LINgreen}{lightgray}}A1 &\cellcolor{\iftoggle{iscolor}{LINgreen}{lightgray}}A1 & \cellcolor{\iftoggle{iscolor}{LINgreen}{lightgray}}A1   \\
                                \hhline{|~|--------|}
                                & SSI Blockchain          & & & & & & &  \\
    \hline 
   \end{tabular}
\end{table}
\Cref{tab:lin:threat-table:data-availability} shows the threat table of our \nth{2} user action, performing registration of data availability.
\begin{table}[t]
  \centering
  \caption{\acl{ld}'s threat table of the \nth{2} user action, performing
            registration of data availability. An \xmark\ in a cell indicates a privacy threat for the
            corresponding threat target. 
            Cells labeled by ``Ax'' are no threats because of the indicated assumptions.}
   \label{tab:lin:threat-table:data-availability}
  \begin{tabular}{ | m{.18\textwidth}<{\centering} | m{.49\textwidth}  || L | L | L | L | L | L | L | }
     \hline
     \multirow{2}{*}{\ac{dfd} Elements} & \multirow{2}{*}{Threat Target} & \multicolumn{7}{c|}{Privacy Threats} \\
                                                                           \cline{3-9}
                                        &                                & \tb{L} & \tb{I} & \tb{N}r & \tb{D}e & \tb{D}i & \tb{U} & \tb{N}c \\
    \hline \hline
    \multirow{2}{*}{Data Store} & Catalog Storage                      & \xmark & \xmark & & & & & \\
                                \hhline{|~|--------} 
                                & Cloud Storage                        & \xmark & \xmark & & & & & \\
    \hline
    \multirow{8}{*}{Info Flow}  & collect authentic data from smart devices    & & & & & & &  \\
                                \hhline{|~|--------}
                                & send encrypted authentic secret-shared data to cloud storage and signed permitted functions & \xmark & \xmark & \xmark & \xmark &\cellcolor{\iftoggle{iscolor}{LINorange}{lightgray}}A3 & &  \\
                                \hhline{|~|--------}
                                & state data availability and specify permitted data analysis    & \xmark & \xmark & \xmark & \xmark &\cellcolor{\iftoggle{iscolor}{LINorange}{lightgray}}A3 & &  \\
                                \hhline{|~|--------}
                                & Update data catalog & & & & & & &  \\
                                \hhline{|~|--------}
                                & Update policies     & & & & & & &  \\
    \hline
    \multirow{2}{*}{Process}    & Backend             & & & & & & &  \\
                                \hhline{|~|--------}
                                & Blockchain nodes    & & & & & & &  \\
    \hline
    \multirow{3}{*}{Entity}     & Data Owner          & & & & & & \xmark &  \\
                                \hhline{|~|--------}
                                & Data Consumer       & & & & & & &  \\
                                \hhline{|~|--------} 
                                & Cloud Storage       & & & & & & &  \\
    \hline 
   \end{tabular}
\end{table}
\Cref{tab:lin:threat-table:data-analysis} shows the threat table of our \nth{3} user action, performing data analysis.
\begin{table}[t]
\centering
   \caption{\acl{ld}'s threat table of the \nth{3} user action, performing data
            analysis. An \xmark\ in a cell indicates a privacy threat for the
            corresponding threat target. 
            Cells labeled by ``Ax'' are no threats because of the indicated assumptions.}
   \label{tab:lin:threat-table:data-analysis}
  \begin{tabular}{ | m{.18\textwidth}<{\centering} | m{.49\textwidth}  || L | L | L | L | L | L | L | }
    \hline
     \multirow{2}{*}{\ac{dfd} Elements} & \multirow{2}{*}{Threat Target} & \multicolumn{7}{c|}{Privacy Threats} \\
                                                                           \cline{3-9}
                                        &                                & \tb{L} & \tb{I} & \tb{N}r & \tb{D}e & \tb{D}i & \tb{U} & \tb{N}c \\
    \hline \hline
    \multirow{3}{*}{Data Store} & Catalog Storage & \xmark & \xmark &   &   &   &   &  \\
                                \hhline{|~|--------} 
                                & Cloud Storage  &   &   & \xmark & \xmark &   &   &  \\
                                \hhline{|~|--------} 
                                & Data Consumer  &   &   &   &   &   &   &  \\
    \hline
    \multirow{3}{*}{Info Flow}  & 4) Request data catalog    & \xmark & \xmark & \xmark & \xmark & \cellcolor{\iftoggle{iscolor}{LINorange}{lightgray}}A3 &   &  \\
                                \hhline{|~|--------} 
                                & 5) Request data analysis   & \xmark & \xmark & \xmark & \xmark & \cellcolor{\iftoggle{iscolor}{LINorange}{lightgray}}A3 &   &  \\
                                \hhline{|~|--------} 
                                & 7) Invoke data analysis    & \xmark & \xmark & \xmark & \xmark & \cellcolor{\iftoggle{iscolor}{LINorange}{lightgray}}A3 &   &  \\
    \hline 
    \multirow{3}{*}{Data Flow}  & 8) Request enc. auth. se-sha. data   & \xmark & \xmark & \xmark & \xmark & \cellcolor{\iftoggle{iscolor}{LINorange}{lightgray}}A3 &   &  \\
                                \hhline{|~|--------} 
                                & 9) Exchange se-sha. data             & \xmark & \xmark & \xmark & \xmark & \cellcolor{\iftoggle{iscolor}{LINorange}{lightgray}}A3 &   &  \\
                                \hhline{|~|--------} 
                                & 10) Return enc. analysis result      & \xmark & \xmark & \xmark & \xmark & \cellcolor{\iftoggle{iscolor}{LINorange}{lightgray}}A3 &   &  \\
    \hline 
    \multirow{3}{*}{Process}    & 6) Check permission of data analysis              &   &   &   &   &   &   &  \\
                                \hhline{|~|--------} 
                                & 9) Check permission \& Perform MPC  
                                     &   &   &   &   & \cellcolor{\iftoggle{iscolor}{LINblue}{lightgray}}A2&   & \cellcolor{\iftoggle{iscolor}{LINblue}{lightgray}}A2 \\
                                \hhline{|~|--------} 
                                & 11) Decrypt analysis result \& Check authenticity &   &   &   &   &   &   &  \\
    \hline 
    \multirow{3}{*}{Entity}     & Cloud Storage  & &  & & & & &  \\
                                \hhline{|~|--------} 
                                & \ac{mpc} Nodes &   &   & & & & &  \\
                                \hhline{|~|--------} 
                                & Data Consumer  & &  & & & & &  \\
    \hline 
   \end{tabular}
\end{table}

\subsection{Threat Elicitation}
After having identified the elements of the DFD susceptible to privacy threats,
the next step is to describe them in detail.
%
In our scenario we make four assumptions, which we present and briefly discuss in the following.

\begin{assumption}\textbf{
\iftoggle{iscolor}{\color{LINgreen}}{}
The \ac{ssi}'s credential issuer is a trusted entity.
}
             If the KRAKEN backend colludes with the credential issuer, the
             KRAKEN backend gets to know not only the \ac{ssi} identity, but also the users'
             real identity. The assumption of an honest \ac{ssi}'s credential issuer, gives
             us the guarantee that the KRAKEN backend cannot link \ac{ssi} identities to real
             identities. 
            
             While this assumption might look overly strong at first glance, it could be achieved by designing the issuer as a distributed party deploying threshold cryptography, or, on a policy-level, through regular audits.
             Furthermore, by informing users about this requirements, they can also check for, e.g., legal relationships between these two entities, before revealing any personal data.
\end{assumption}

\begin{assumption}\textbf{
\iftoggle{iscolor}{\color{LINblue}}{}
At least one \ac{mpc} node is honest.
}
For \ac{mpc}, we can use protocols which give security guarantees
although all parties bar one are malicious. We refer to these kind of protocols
as \textit{fully-malicious protocols}. Thus, with this assumption and the
respective protocols, no \ac{mpc} node either gets to know the underlying data,
nor the analysis result. Furthermore, due to this assumption and the provided
policies by the data owners, we do not need to worry about the case, that
\ac{mpc} nodes compute without authorization, since the honest \ac{mpc} node
aborts the computation, which leads to a global termination of the computation.           

To minimize the impact of this assumption, the hosts of the MPC nodes should be
carefully selected in order to minimize intended collusions among them or their
simultaneous corruptions. A possible approach might be to include more than the
required three computation nodes in the ecosystem, and letting the user choose
which nodes to support when encrypting her data. Note that a selected
node can still get corrupted, but it might be less likely. Though, after all,
(1) the probability that all nodes get corrupted might decrease too, and (2)
all nodes need to get corrupted to leak sensitive data when using a
fully-malicious protocol.

Especially for highly sensitive data, one of these nodes might even be deployed at the data owners.
Thereby, the honest node is in control of the data owner and hence leaking sensitive data is low.
\end{assumption}

\begin{assumption}\textbf{
\iftoggle{iscolor}{\color{LINorange}}{}
Each communication between actors of different trust domains is secured using \ac{tls}.
}
             When two actors of different trust domains communicate, this assumption guarantees that no content is leaked during
             transit. Only the communication's metadata can still leak, like the
             IP addresses of the source and target.
             Secure communication protocols are heavily deployed in many scenarios, and can be considered state-of-the-art.
\end{assumption}

\begin{assumption}\textbf{Anybody who gains access to any of trust boundaries is considered to
          have the same possibilities as the corrupted entity itself, and trust boundaries are implemented
          in a secure way.}
             If someone hacks into a any of trust domains, the intruder (normally) gets access to the corresponding data stores,
             processes, and data flow. Hence the intruder has full control over the
             trust domain's system.

            To a major extent, this is rather a simplification than an assumption, as it significantly simplifies the analysis.
            We do not need to distinguish between insider attacks (e.g., a system administrator), and an external attack (e.g., an attacker partial gaining control over an entity), but we assume that once a trust boundary is violated, the entire entity is fully controlled by the adversary.
\end{assumption}


\subsubsection{Mapping \acl{ld}'s Privacy Threats to the \acp{dfd}.}

In this section the threats identified using the threat tables are described in detail.

\begin{threat}[Linkability in one or more storages] \label{threat:Lstorage}
An insider of KRAKEN links data coming from the catalog, credentials, policies or purchases storages.
\begin{threatdesc}
  \item[Assets, stakeholder, threats:]
Linking different users or different information of the same user could lead to gain more information
about users than expected.
\item[Primary misactor:]
An internal user that has access to the data storages of the backend and/or of the internal blockchain.
\item[Basic flow:]
\begin{enumerate*}
  \item The insider gains specific information by querying the data store.
  \item The obtained set of information can be linked.
\end{enumerate*}
\item[Preconditions:]
The user has updated the system with some informations or is at least registered.
\item[DFD elements:]
Credentials storage, Catalog storage, Policies storage, Purchases storage, Cloud storage.
\item[Remarks:]
  This threat could lead to identification.
  When applied to the credentials storage, the probability is much lower as credentials have a high level of minimization of information.
\end{threatdesc}
\end{threat}

\begin{threat}[Identifiability in one or more storages] \label{threat:Istorage}
An insider of KRAKEN identifies one or more users in a set of data coming from one or more storages.
\begin{threatdesc}
\item[Assets, stakeholder, threats:]
The identity of the user must be unknown in the KRAKEN.
\item[Primary misactor:]
An internal user that has access to the data storages of the backend and/or of the internal blockchain.
\item[Basic flow:]
\begin{enumerate*}
  \item The insider gains specific information by querying one or more data stores.
  \item The obtained set of information can be linked and can lead to identification of one or more users.
\end{enumerate*}
\item[Preconditions:]
The user has updated the system with some information or is at least registered.
\item[DFD elements:]
Credentials storage, Catalog storage, Policies storage, Purchases storage, Cloud storage.
\end{threatdesc}
\end{threat}

\begin{threat}[Detectability of data existence] \label{threat:Dexistance}
The user uploads the data on the cloud without publishing on KRAKEN, revealing the existence of data.
\begin{threatdesc}
\item[Assets, stakeholder, threats:]
The detection of the existence of the data must take place at the will of the user.
\item[Primary misactor:]
The cloud or an external actor.
\item[Basic flow:]
  The misactor checks periodically the cloud storage until the data is uploaded.
\item[DFD elements:]
Cloud storage
\end{threatdesc}
\end{threat}

\begin{threat}[Detectability in communication between different trust domains] \label{threat:Dcommunication}
An internal/external actor can detect user actions by listening to requests.
\begin{threatdesc}
\item[Assets, stakeholder, threats:]
The detectability of user actions is not expected outside of the scope of the interested actors.
\item[Primary misactor:]
A skilled internal/external actor that has access to the network of the user and can inspect user's packets.
\item[Basic flow:]
\begin{enumerate*}
  \item The misactor intercepts packets between a user and KRAKEN.
  \item Whenever a packet is sent, an action has been detected.
\end{enumerate*}
\item[DFD elements:]
All the data flows between two different trust domains.
\item[Remarks:]
  This threat disclosure of information is not expected as the communication happens through TLS.
\end{threatdesc}
\end{threat}

\begin{threat}[Linkability of IP addresses in communication between different trust domains]
An internal/external actor can link different events to the same user by listening to user's requests.
  \begin{threatdesc}
\item[Assets, stakeholder, threats:]
Any information that can be gained by linking user actions are not expected to be known by anyone except the user.
\item[Primary misactor:]
A skilled internal/external actor that has access to the network of the user and can inspect user's packets.
\item[Basic flow:]
\begin{enumerate*}
  \item The misactor intercepts packets between a user and KRAKEN.
  \item Whenever a packet is sent, IP addresses are collected.
  \item The misactor links packets with the same IP.
\end{enumerate*}
\item[DFD elements:]
All the data flows between two different trust domains.
\item[Remarks:]
  This threat disclosure of information is not expected as the communication happens through TLS.
\end{threatdesc}
\end{threat}

\begin{threat}[Linkability of IP addresses in communication between different trust domains leads to identifiability]
An internal/external actor can identify users by linking different events to the same IP by listening to user's requests.
\begin{threatdesc}
\item[Assets, stakeholder, threats:]
User's identity and any information that can be gained by linking user actions are not expected to be known by anyone except the user.
\item[Primary misactor:]
A skilled internal/external user that has access to the network of the user and can inspect user's packets and knows or can link
to an IP address the user's identity.
\item[Basic flow:]
\begin{enumerate*}
  \item The misactor intercepts packets exchanged between a user and KRAKEN.
  \item Whenever a packet is sent, IP addresses are collected.
  \item The misactor links packets with the same IP.
  \item The gained information, together with any information that can link the IP to a user (e.g., insecure traffic with other systems) leads to the identification of the user.
\end{enumerate*}
\item[DFD elements:]
All the data flows between two different trust domains.
\end{threatdesc}
\end{threat}

\begin{threat}[Non-repudiation of encrypted data] \label{Nr-UA3-1-ED} 
The cloud storage cannot repudiate that encrypted data is available.
\begin{threatdesc}
\item[Primary misactor:]
Data stores which do not handle data access properly.
\item[DFD elements:]
Cloud storage (data store; \ac{ua} 2/3).
\end{threatdesc}
\end{threat}

\begin{threat}[Non-repudiation of communication between different trust domains] \label{Nr-UA3-2-C} 
An entity cannot repudiate that he sent a message to another entity within a
different trust domain.
\begin{threatdesc}
\item[Primary misactor:]
An external user that has access to the network of the user and can inspect
user's packets.
\item[DFD elements:]
All data flows between two different trust domains.
\end{threatdesc}
\end{threat}



\begin{threat}[Unawareness of the data owner] \label{U-1-DO} 
First, a data owner provides data for which he is not
allowed, such as by national law. Second, a data owner does not take care of the
defined analysis policies/permissions, such that a consumer could learn
something about the owner based on the analysis result. For example, if an owner
allows an analysis without any other owners in addition (aggregated analysis),
then, e.g.,  an average would reveal the actual data.
  \begin{threatdesc}
\item[Primary misactor:]
A data owner making data available.
\item[DFD elements:]
Data owner (entity; \ac{ua} 2).
  \end{threatdesc}
\end{threat}

\begin{threat}[Non-deletion of data in cloud storage] \label{U-2-DO} 
  The data owner is not aware that the cloud storage is in possession of his data.
    \begin{threatdesc}
  \item[Primary misactor:]
  A cloud storage not deleting user's data.
  \item[Basic flow]
  \begin{enumerate*}
   \item The data owner requests the cloud storage to delete his data.
   \item The cloud storage doesn't delete the data.
   \item The data owner is not aware that the data is stored on the cloud storage.
  \end{enumerate*}
  \item[DFD elements:]
  Data owner (entity; \ac{ua} 2).
    \end{threatdesc}
  \end{threat}


\subsection{Prioritizing Threats}

For the prioritization of the threats, first a likelihood and impact value is assigned to every threat identified in the threat table.
Both values are taken from low, medium, and high indicating a low to high likelihood and impact, respectively.
The likelihood value depends on the joint evaluation of difficulty and outcome of performing the specific action, while the impact
value depends on threatened assets where identifiability and disclosure of information are high impact, linkability is medium and 
Non-repudiation, Detectability, Unawareness, Non-compliance are low.
\Cref{tab:lin:threat-table:priorities} shows how threats are prioritized depending on likelihood and impact values.

\begin{table}[t!]
\centering
\caption{Threat prioritization depending on likelihood and impact.}
   \label{tab:lin:threat-table:priorities}
\scalebox{0.92}{
\begin{tabular}{|c|c|c|}
\hline
Likelihood & Impact & Priority \\
\hline \hline
low & low & \multirow{3}{*}{low} \\
low & medium & \\
medium & low & \\
\hline
\end{tabular}
\quad
\begin{tabular}{|c|c|c|}
\hline
Likelihood & Impact & Priority \\
\hline \hline
low & high & \multirow{3}{*}{medium}\\
medium & medium & \\
high & low & \\
\hline
\end{tabular}
\quad
\begin{tabular}{|c|c|c|}
\hline
Likelihood & Impact & Priority \\
\hline \hline
medium & high & \multirow{3}{*}{high} \\
high & medium &  \\
high & high &  \\
\hline
\end{tabular}
}
\end{table}

\begin{table}[t]
\centering
   \caption{Overview of threat prioritization. Threats that are not effective due to our assumptions are not included in the table.}
   \label{tab:lin:threat-table:prioritization}
  \begin{tabular}{ | m{.6\textwidth}<{\raggedleft} || c | c || c | }
    \hline
      \multicolumn{1}{|c||}{Threat} & Likelihood & Impact & Priority \\
    \hline \hline
      Linkability in one or more storages
        & medium & medium & medium \\
    \hline
      Identifiability in one or more storages
        & low & high & medium \\
    \hline
      Detectability of data existence
        & medium & low & low \\
    \hline
      Detectability in communication between different trust domains
        & low & low & low \\
    \hline
      Linkability of IP addresses in communication between different trust domains
        & low & medium & low \\
    \hline
      Linkability of IP addresses in communication between different trust domains leads to identifiability
        & low & high & medium\\
    \hline 
        Non-repudiation of encrypted data 
        & low & low & low \\
    \hline 
        Non-repudiation of communication between different trust domains  
        & low & low & low \\
    \hline
      Unawareness of the data owner
        & low & high & medium \\
    \hline
    Non deletion of data in cloud storage
        & low & low & low \\
    \hline 
   \end{tabular}
\end{table}

\Cref{tab:lin:threat-table:prioritization} gives an overview of the prioritization of the identified threats.
In the following, we give a brief justification for each threat.

\begin{description}[style=sameline]
  \item[Linkability in one or more storages.] In this threat the likelihood value is medium as even if the misactor needs to be an insider, exploiting more than one storages leads to better outcomes in trying to link user's data.
  The impact is medium as the threatened asset is the linkability of user's data, that if combined with identifiability reveals which users performed certain actions.

  \item[Identifiability in one or more storages.] The likelihood value is low as the misactor would need more information other than the ones contained in the KRAKEN system to identify one or more users.
  The impact is high as the threatened asset is the identity of users that is considered high priority asset.

  \item[Detectability of data existence.] The likelihood is medium as the threatened information is public by default. The misactor could be an external user without any specific capability that needs to know by other means that the specific data is destined to KRAKEN.
  In the case where the misactor is the cloud storage that may know the identity of the user, the cloud storage would still need to know by other means that the specific data is destined to KRAKEN.
  In a hospital scenario, If a patient decides to adopt the hospital's cloud system, the hospital could
  make assumptions on the content of the dataset by linking the detection of the dataset existence with information related to
  the patient. However, this situation is highly unlikely as the user can choose any cloud system
  without relying on the hospital's one.
  The impact is low as the data is always encrypted, existence of data may be detected, but the data itself does not leak.

  \item[Detectability in communication between different
    trust domains.] The likelihood value is low as the misactor is an external
    skilled individual that has access to the network of the user or to the KRAKEN
    network. The impact is low as the threatened asset is the detectability of user
    actions, which is considered a low-priority asset.

  \item[Linkability of IP addresses in communication between different trust domains.] This threat depends on the same actions and actor
  needed to perform the previous one, so the likelihood is the same.
  The impact is medium as the threatened asset is the linkability of user's data, that if combined with identifiability reveals which users performed certain actions.

  \item[Linkability of IP addresses in communication between different trust domains leads to identifiability.] This threat depends on the same actions and actor
  needed to perform the previous one, so the likelihood is the same.
  The impact is high as the threatened asset is the identity of users that is considered high priority asset.

  \item[Non-repudiation of encrypted data.]
    As cloud-storage providers usually use unguessable file links, the
    likelihood for this threat is low.
    The impact is low as one cannot identify the receiver of the ciphertext recover its content.

  \item[Non-repudiation of communication between different trust domains.]
    Similar as for detectability of communication, likelihood and impact are low.

  \item[Unawareness of the data owner.] The likelihood value is low as the personal data provided belongs to the user and therefore it is her own interest to provide data
  that does not affect her in terms of non compliance with regulations.
  Moreover (for the second case) the outcome of publishing the analysis of a dataset without a pool of other user's datasets would not be appealing for a possible buyer.
  The impact is high as the threatened asset is the personal information of users that is considered high priority asset.

  \item[Non deletion of data in cloud storage.] The likelihood value is low as the outcome of performing this action would lead the cloud storage to have an encrypted dataset that is not
  possible to consume in any way. Because of Assumption 2, the cloud storage cannot collaborate with the MPC nodes to unveil the data as at least one of them is honest.
  The impact is low as the threatened asset is the unawareness of users that is considered low priority.
\end{description}


\subsection{Mitigating Threats}
For every threat in non low priority, we propose a set of mitigations expressed in the
following list:
\begin{description}[style=sameline]
  \item[Linkability in one or more storages.]
        To mitigate the threat on the SSI storage side, on registration phase the system can request to the user
        the minimum set of credentials required to allow the user to get registered and do not lead to linkability/identification.
        To mitigate the threat on the other storages, the system can display a suggestion to user saying to non include any
        identifiable information before the publication of any product.
  \item[Identifiability in one or more storages.]
        This threat depends on the previously described threat ``Linkability in one or more storages'', the mitigation
        applied in that threat mitigate consequently also this one.
  \item[Linkability of IP addresses in communication between different trust domains leads to identifiability.]
        To avoid the misactor to understand that the communication is happening with KRAKEN, avoiding linkability and
        resulting identifiability, we propose onion routing like Tor~\cite{DBLP:conf/uss/DingledineMS04}.
  \item[Unawareness of the data owner.]
        The mitigation can be implemented on the user's frontend side in two complementing ways.
        First, the system provides thorough documentation that explains potential risks when offering certain data sets for data analytics.
        Second, based on the type of data and the acceptable function families, privacy metrics~\cite{DBLP:journals/csur/WagnerE18} are displayed to make the user aware of any risks.
        Thereby, the system is able to warn the user, e.g., before allowing the computation of an average but where the user's input is the only considered data set.
\end{description}


\subsection{Privacy Analysis Outcome}
We adopted an iterative approach in the design of the architecture that used the LINDDUN privacy analysis
to identify threats and plan the changes for the iterations.
It's worth mentioning the most relevant changes that this approach generated.
The DFDs (\Cref{fig:lind:dfd:user-reg,fig:lind:dfd:data-analysis}) highlight the differences in information and data flow.
This division in typology of data flows has been key leading us to construct an architecture where personal data is exchanged solely between data owner and consumer, without passing through centralized parties unencrypted.

Another key element derived from the analysis is the use of group signatures. Public keys of the users represent a risk for identification of the user or
could be linked to other actions. For this reason we decided to adopt group signatures to allow the user to sign data and permitted functions on behalf of a group.
In this way the user can demonstrate to be part of the users of KRAKEN and can sign data and functions anonymously while retaining authenticity.

Finally, we identified a set of threats that do not imply architectural changes, but instead have to be considered in a development context.
These threats and their mitigations affect single elements of the
architecture: the Credentials storage,
Catalog storage, Policies storage, Purchases storage, Cloud storage, and the Data
owner. A set of changes need to be implemented in these elements to address
the mitigations. In particular, we applied a principle of data
minimization in the context of the backend and the blockchain storages, while in
the user software development, we considered a set of tools to be provided to
the Data owner for documentation, analytics, and threat detection purposes.

\section{Conclusion and Future Work}\label{sec:conclusion}
In this paper, we presented a privacy-preserving data market platform and analysed its privacy-guarantees following the LINDDUN methodology.
The proposed solution allows users to sell data without any disclosure of
information in regards to the data itself.  The LINDDUN analysis revealed
threats related to linkability and disclosure of information that could have a
relevant impact, however these threats have a low likelihood and the system's
methodologies in collecting information related to those threats is implemented
in a way that highly minimizes the collected information.  The LINDDUN analysis
does not reveal any threat related to disclosure of information of the owners
data sets.

The proposed architecture is based on cryptographic mechanisms, and in
particular \acf{mpc}. With that approach, a data consumer is able to obtain
privacy-preserving data analysis results from data owners, while the consumer
receives only the analysis result. Furthermore, our marketplace does neither
learn the owners' data content nor the analysis result, but only metadata,. As
opposed to Agora~\cite{DBLP:journals/iacr/KoutsosPCTH20}, where the broker, their
market place, gets to know the analysis result. Our security guarantees in terms
of privacy analysis, however, depend on the assumption that at least one MPC
node behaves honestly. A possible future work would be on realising a trust
measurement to drive the choice of MPC nodes.  Another possible field of
research would go towards moving the MPC computations on data owners. The
obstacle to overcome in this case would be the problem of user availability
during the computation, as users typically do not have constantly running
servers available.

The architecture (cf. \Cref{sec:architecture}) only considers lump sums to reimburse the data owner.
However, it might be practically more preferable to get paid \emph{per usage}, i.e., whenever one's data is actually used in a computation, resulting in additional privacy challenges.
A straightforward solution might be to add an exchange service.
This service would then be able to link usages of a user's data, thereby being able to profile a user, especially when collaborating with a data consumer. 
Thus, such a service would need to be highly trusted akin to traditional banks in a physical world.
An alternative approach could be to leverage privacy-friendly crypto currencies like Monero~\cite{DBLP:journals/ledger/NoetherM16} or z.cash~\cite{DBLP:conf/sp/MiersG0R13}.
To further increase trust in the system, the \ac{mpc} nodes could publish cryptographic yet privacy-preserving proofs which data was used for which computation, such that a user could audit that she was indeed paid for every computation involving her data.
The precise format of such auxiliary outputs of the \ac{mpc} nodes is currently being investigated.




\bibliographystyle{splncs04.bst}
\bibliography{dblp,bib}

\begin{thebibliography}{10}
\providecommand{\url}[1]{\texttt{#1}}
\providecommand{\urlprefix}{URL }
\providecommand{\doi}[1]{https://doi.org/#1}

\bibitem{medicalchain}
Medicalchain: Whitepaper 2.1 (2018),
  \url{https://medicalchain.com/Medicalchain-Whitepaper-EN.pdf}

\bibitem{enveil}
{Enveil: Encrypted Veil} (2020), \url{https://www.enveil.com/}

\bibitem{www:insurance-personal-data}
Allen, M.: {Health Insurers Are Vacuuming Up Details About You - And It Could
  Raise Your Rates} (2020),
  \url{https://www.propublica.org/article/health-insurers-are-vacuuming-up-details-about-you-and-it-could-raise-your-rates}

\bibitem{www:apple-health}
Apple-Inc.: {A more personal Health app. For a more informed you} (2020),
  \url{https://www.apple.com/ios/health/}

\bibitem{DBLP:conf/innovations/BitanskyCCT12}
Bitansky, N., Canetti, R., Chiesa, A., Tromer, E.: From extractable collision
  resistance to succinct non-interactive arguments of knowledge, and back
  again. In: ITCS. pp. 326--349. ACM (2012)

\bibitem{DBLP:journals/ijisec/BogdanovNTW12}
Bogdanov, D., Niitsoo, M., Toft, T., Willemson, J.: High-performance secure
  multi-party computation for data mining applications. Int. J. Inf. Sec.
  \textbf{11}(6),  403--418 (2012)

\bibitem{DBLP:journals/cacm/BonehSW12}
Boneh, D., Sahai, A., Waters, B.: Functional encryption: a new vision for
  public-key cryptography. Commun. {ACM}  \textbf{55}(11),  56--64 (2012)

\bibitem{DBLP:conf/socialcom/BrickellL10}
Brickell, E., Li, J.: Enhanced privacy {ID} from bilinear pairing for hardware
  authentication and attestation. In: SocialCom/PASSAT. pp. 768--775. {IEEE}
  (2010)

\bibitem{fix:journals/iacr/HelmingerKRW20}
Bruni, A., Helminger, L., Kales, D., Rechberger, C., Walch, R.: {Privately
  Connecting Mobility to Infectious Diseases via Applied Cryptography}. {IACR}
  Cryptol. ePrint Arch.  \textbf{2020}, ~522 (2020)

\bibitem{DBLP:conf/scn/CamenischHL06}
Camenisch, J., Hohenberger, S., Lysyanskaya, A.: Balancing accountability and
  privacy using e-cash (extended abstract). In: SCN. LNCS, vol.~4116, pp.
  141--155. Springer (2006)

\bibitem{DBLP:conf/eurocrypt/CamenischL01}
Camenisch, J., Lysyanskaya, A.: An efficient system for non-transferable
  anonymous credentials with optional anonymity revocation. In: EUROCRYPT.
  LNCS, vol.~2045, pp. 93--118. Springer (2001)

\bibitem{www:data-trend}
Chandler, S.: {We're giving away more personal data than ever, despite growing
  risks} (2020),
  \url{https://venturebeat.com/2019/02/24/were-giving-away-more-personal-data-than-ever-despite-growing-risks/}

\bibitem{DBLP:conf/crypto/Chaum82}
Chaum, D.: Blind signatures for untraceable payments. In: CRYPTO. pp. 199--203.
  Plenum Press, New York (1982)

\bibitem{DBLP:conf/eurocrypt/ChaumH91}
Chaum, D., van Heyst, E.: Group signatures. In: EUROCRYPT. LNCS, vol.~547, pp.
  257--265. Springer (1991)

\bibitem{sharemind}
Cybernetica: {Sharemind MPC}. \url{https://sharemind.cyber.ee/sharemind-mpc/}
  (2020)

\bibitem{DBLP:journals/re/DengWSPJ11}
Deng, M., Wuyts, K., Scandariato, R., Preneel, B., Joosen, W.: A privacy threat
  analysis framework: supporting the elicitation and fulfillment of privacy
  requirements. Requir. Eng.  \textbf{16}(1),  3--32 (2011)

\bibitem{DBLP:journals/corr/abs-1712-01767}
Der, U., J{\"{a}}hnichen, S., S{\"{u}}rmeli, J.: Self-sovereign identity -
  opportunities and challenges for the digital revolution. CoRR
  \textbf{abs/1712.01767} (2017)

\bibitem{DBLP:conf/uss/DingledineMS04}
Dingledine, R., Mathewson, N., Syverson, P.F.: Tor: The second-generation onion
  router. In: {USENIX}. pp. 303--320. USENIX (2004)

\bibitem{duality}
{Duality Technologies Inc}: {Duality}. \url{https://dualitytech.com/} (2020)

\bibitem{DBLP:journals/corr/abs-2001-08832}
Fernandez, D., Futoransky, A., Ajzenman, G., Travizano, M., Sarraute, C.:
  Wibson protocol for secure data exchange and batch payments. CoRR
  \textbf{abs/2001.08832} (2020)

\bibitem{www:garmin}
Garmin-Ltd.: {connect: Fitness at your fingertips} (2020),
  \url{https://connect.garmin.com/}

\bibitem{DBLP:conf/stoc/GoldwasserMR85}
Goldwasser, S., Micali, S., Rackoff, C.: The knowledge complexity of
  interactive proof-systems (extended abstract). In: STOC. pp. 291--304. ACM
  (1985)

\bibitem{www:google-fit}
Google: {Google Fit: Coaching you to a healthier and more active life} (2020),
  \url{https://www.google.com/fit/}

\bibitem{DBLP:conf/acns/Groth05}
Groth, J.: Non-interactive zero-knowledge arguments for voting. In: ACNS. LNCS,
  vol.~3531, pp. 467--482 (2005)

\bibitem{DBLP:journals/js/GusevPSKKSASTT20}
Gusev, M., Poposka, L., Spasevski, G., Kostoska, M., Koteska, B., Simjanoska,
  M., Ackovska, N., Stojmenski, A., Tasic, J.F., Trontelj, J.: Noninvasive
  glucose measurement using machine learning and neural network methods and
  correlation with heart rate variability. J. Sensors  \textbf{2020},
  9628281:1--9628281:13 (2020)

\bibitem{DBLP:journals/iacr/IonKNPSSSY17}
Ion, M., Kreuter, B., Nergiz, E., Patel, S., Saxena, S., Seth, K., Shanahan,
  D., Yung, M.: Private intersection-sum protocol with applications to
  attributing aggregate ad conversions. {IACR} Cryptol. ePrint Arch.
  \textbf{2017}, ~738 (2017)

\bibitem{DBLP:journals/iacr/KimLAP20}
Kim, H., Lee, Y., Abdalla, M., Park, J.H.: Practical dynamic group signature
  with efficient concurrent joins and batch verifications. {IACR} Cryptol.
  ePrint Arch.  \textbf{2020}, ~921 (2020)

\bibitem{DBLP:journals/iacr/KoutsosPCTH20}
Koutsos, V., Papadopoulos, D., Chatzopoulos, D., Tarkoma, S., Hui, P.: Agora:
  {A} privacy-aware data marketplace. {IACR} Cryptol. ePrint Arch.
  \textbf{2020}, ~865 (2020)

\bibitem{www:kraken-h2020}
{KRAKEN Consortium}: {The Project | KRAKEN} (2020),
  \url{https://www.krakenh2020.eu/the_project/overview}

\bibitem{www:linddun}
linddun.org: {LINDDUN privacy engineering} (2020),
  \url{https://www.linddun.org/}

\bibitem{www:forbes-data-generation-trend}
Marr, B.: {How Much Data Do We Create Every Day? The Mind-Blowing Stats
  Everyone Should Read} (2020),
  \url{https://www.forbes.com/sites/bernardmarr/2018/05/21/how-much-data-do-we-create-every-day-the-mind-blowing-stats-everyone-should-read/}

\bibitem{DBLP:conf/sp/MiersG0R13}
Miers, I., Garman, C., Green, M., Rubin, A.D.: Zerocoin: Anonymous distributed
  e-cash from bitcoin. In: {IEEE} S\&P. pp. 397--411. {IEEE} (2013)

\bibitem{DBLP:conf/edbt/Morley-Fletcher17}
Morley{-}Fletcher, E.: {MHMD:} my health, my data. In: {EDBT/ICDT} Workshops.
  {CEUR} Workshop Proceedings, vol.~1810. CEUR-WS.org (2017)

\bibitem{DBLP:journals/csr/MuhleGGM18}
M{\"{u}}hle, A., Gr{\"{u}}ner, A., Gayvoronskaya, T., Meinel, C.: A survey on
  essential components of a self-sovereign identity. Comput. Sci. Rev.
  \textbf{30},  80--86 (2018)

\bibitem{www:fitbit-for-health}
Muoio, D.: {Fitbit launches large-scale health study to detect a-fib via heart
  rate sensors, algorithm} (2020),
  \url{https://www.mobihealthnews.com/news/fitbit-launches-large-scale-consumer-health-study-detect-fib-heart-rate-sensors-algorithm}

\bibitem{www:google-covid-tracking}
Muoio, D.: {Google mobilizes location tracking data to help public health
  experts monitor COVID-19 spread} (2020),
  \url{https://www.mobihealthnews.com/news/google-mobilizes-location-tracking-data-help-public-health-experts-monitor-covid-19-spread}

\bibitem{DBLP:journals/ledger/NoetherM16}
Noether, S., Mackenzie, A.: Ring confidential transactions. Ledger  \textbf{1},
   1--18 (2016)

\bibitem{DBLP:journals/cacm/Shamir79}
Shamir, A.: How to share a secret. Commun. {ACM}  \textbf{22}(11),  612--613
  (1979)

\bibitem{todd2017towards}
Todd, C., Salvetti, P., Naylor, K., Albatat, M.: {Towards non-invasive
  extraction and determination of blood glucose levels}. Bioengineering
  \textbf{4}(4), ~82 (2017)

\bibitem{DBLP:journals/csur/WagnerE18}
Wagner, I., Eckhoff, D.: Technical privacy metrics: {A} systematic survey.
  {ACM} Comput. Surv.  \textbf{51}(3),  57:1--57:38 (2018)

\bibitem{DBLP:conf/focs/Yao82b}
Yao, A.C.: Protocols for secure computations (extended abstract). In: FOCS. pp.
  160--164. {IEEE} (1982)

\end{thebibliography}

\end{document}